\documentclass[11pt]{article}
\usepackage[top=1.00in, bottom=1.0in, left=1in, right=1in]{geometry}

\usepackage{graphicx}
\usepackage{natbib}
\usepackage{amsmath}
\usepackage{xr-hyper}
\usepackage{hyperref}

\usepackage{fancyhdr}
\begin{document}

\title{A four-step Bayesian workflow for improving ecological science}
\author{EM Wolkovich$^{1*}$, T Jonathan Davies$^{1,2}$, William D Pearse$^{3,4}$ \& Michael Betancourt$^{5}$}
\maketitle

\noindent $^{1}$ Forest and Conservation Sciences, University of British Columbia, Vancouver, BC V6T 1Z4, Canada\\
$^{2}$ Botany, University of British Columbia, Vancouver, BC V6T 1Z4, Canada\\
$^{3}$ Department of Life Sciences, Imperial College London, Ascot SL5 7PY, United Kingdom \\
$^{4}$ Alan Turing Institute, British Library, 96 Euston Road, London NW1 2DB, United Kingdom \\
$^{5}$ Symplectomorphic, LLC, New York, NY 10026, USA \\
$^{*}$ \url{mailto: e.wolkovich@ubc.ca}

\vspace{3ex}
\vspace{3ex}

\noindent \emph{Data availability:} No new data, but full code and raw data are provided in the supplement for an example workflow.

\newpage

\abstract{Growing anthropogenic pressures have increased the need for robust predictive models. Meeting this demand requires approaches that can handle bigger data to yield forecasts that capture the variability and underlying uncertainty of ecological systems. Bayesian models are especially adept at this and are growing in use in ecology. Yet many ecologists today are not trained to take advantage of the bigger ecological data needed to generate more flexible robust models. Here we describe a broadly generalizable workflow for statistical analyses and show how it can enhance training in ecology. Building on the increasingly computational toolkit of many ecologists, this approach leverages simulation to integrate model building and testing for empirical data more fully with ecological theory.  In turn this workflow can fit models that are more robust and well-suited to provide new ecological insights---allowing us to refine where to put resources for better estimates, better models, and better forecasts.}

\setlength{\parindent}{0pt}
\setlength{\parskip}{7pt}

\section*{Introduction}
In recent years, as ecologists have worked to develop global predictive models, they have developed ever larger datasets \citep{Hampton2013}. These bigger data, however, are also messier data. Such data generally requires a model of both the underlying biological processes and how the measurements were made. Some fields have long used these types of models \citep[generally in fields focused on inferring population sizes of things people want to eat or manage,][]{muthuku2008,zheng2007,trijoulet2018,strinella2020potential}. Most, however, have not. This has left many researchers to try to adapt what they were trained in---traditional statistical methods (e.g. $F$ and $t$ tests) and a strong focus on null hypothesis testing---to increasingly complex datasets. Often they have done this by fitting multi-way interaction terms, using random effects to correct for group level factors, or comparing across a large suite of models. 

Common embedded statistical approaches do not often align with ecology's aims today. Beyond the reality that most traditional methods are fragile when used beyond the cleaner, simpler experiments these methods assume (e.g. spatial, temporal and phylogenetic correlations often violate common independence assumptions), they will usually fail to produce robust, reproducible results. Multi-way interaction terms can make main effects hard to interpret and require much larger sample sizes to estimate reliably. Null hypotheses are rarely true and can lead to confusion over what is scientifically important versus `significant' \citep{gelmanhill,muff2022rewriting}. Many model comparison and machine learning approaches prefer models whose inferences match the idiosyncrasies and biases in the available data, but don’t generalize. The disconnect between the traditional statistical tools available in ecology and the field's fundamental theory is being made ever-clearer as the need for robust ecological forecasts grows. 

Bayesian approaches provide a pathway to build powerful models that can transform how we understand our systems as largescale ecological data become increasingly available. Recognizing this, many in ecology are increasingly using Bayesian methods \citep{anderson2021trends}. New algorithms \citep[e.g. Hamiltonian Monte Carlo,][]{nuts2014,betan2019} that have made fitting and implementing Bayesian models faster, more robust and---in many ways---easier \citep{Carpenter:2017stan}. Fitting larger and sometimes more complex models, however, presents challenges that are frequently not addressed in traditional ecological training. We suggest that many of these challenges can be overcome by approaching analyses through specific workflows \citep{betanworkflow,grinsztajn2021,vandeschoot2021}, which themselves are built on a process of how to do not just statistics, but how to do science \citep{box1976science}. 

Such approaches move away from a focus on null hypothesis testing, towards estimating effect sizes, using models calibrated (see Table \ref{tab:glossary}) and better understood through simulating data at multiple steps---using a number of skills often reserved in ecology more for `theorists' than empirical ecologists.  But this theoretical-vs-empirical divide ignores that the average modern ecologist is computational, and thus already has many of the basic skills to build bespoke models. 

Here we provide a simplified---but powerful---workflow that builds on new insights from statistics  \citep{betanworkflow,vandeschoot2021} and the increasingly computational nature of ecology today. By integrating bespoke model building more fully with ecological theory and understanding---and vice versa---this approach can fit models that are more robust and better-suited to providing new ecological insights and improved predictions. We introduce our workflow assuming a Bayesian statistical framework; however, it can also be applied to other statistical inference methods. 

\section*{A basic Bayesian workflow}
Statistical analyses are designed for inference---to learn about some process, effect or behavior from data. Robust analyses yield inferences consistent with the underlying truth more often than not (quantifying this consistency is called calibration, and is a critical part of using models for inference, see Table \ref{tab:glossary}). Because we do not know the `truth' how we approach our analyses---and our inference---is a critical component of how we do science. Fragile statistical analyses can lead science away from the type of repeatable and generalizable findings expected when inferences align with the truth. For example, an overly zealous focus on $p$-values has led to a replication crises in several fields, where results seem most likely the outcome of noisy data combined with a search for statistical significance through many models \citep[effectively a garden of forking paths,][]{halsey2015,loken2017}. Some model selection approaches, including new machine learning methods, try to avoid this by comparing across models, but may not generalize to provide useful forecasts.  This is especially true when forecasts have to adapt to changes in the underlying ecology.

We argue that robust analyses comes from explicitly building and challenging a model (or set of models) in an organized sequence of steps---a workflow. We outline one such workflow using a Bayesian approach (with an example in the supplement shown in \textsf{R} and \textsf{Stan}). 

In ecology, Bayesian methods often shift focus away from null hypothesis testing and point estimates and towards evaluating more fully a model's output, including uncertainty---often using a specific model designed for the system and question at hand. These bespoke models can be fit by applying Bayes' theorem, which generates a $posterior$ distribution from a combination of a $likelihood$ and a $prior$ distribution (an initial uncertainty estimate derived from basic ecological knowledge), and using iterative algorithms (e.g. MCMC, Markov Chain Monte Carlo) that provide samples that can be used to extract information from the posterior distribution (for more, see \emph{A brief review of statistical inference using Bayesian approaches} in the Supplement).

\begin{figure}[ht]
\centering
\noindent \includegraphics[width=1\textwidth]{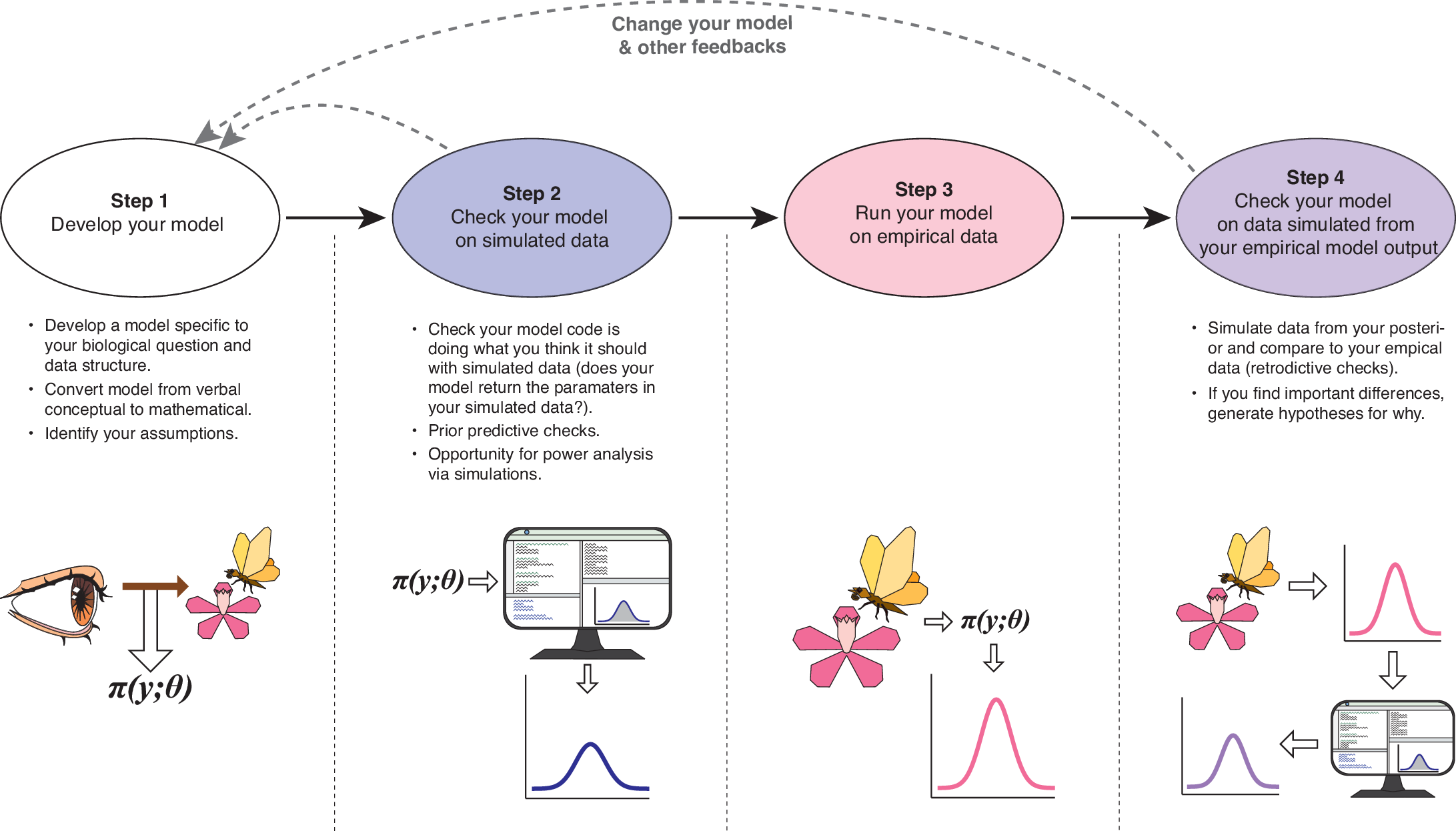}
\caption{The four-step iterative workflow we outline can help design models for specific ecological questions, data and aims---which makes this a statistical workflow that can naturally become a scientific workflow. It makes the step that many of us focus on---running your model on your empirical data (Step 3)---far more straightforward and insightful by using simulations both before (Step 2) and after (Step 4) it to better understand the model and data together.}
\label{fig:workflow}
\end{figure}

Our workflow describes what we consider the major steps for Bayesian model fitting (Fig. \ref{fig:workflow}). This includes model development (Step 1) and calibration (Step 2), inference (Step 3), and then model improvement (Step 4). Many of these steps will be familiar to statistical ecologists, but are often overlooked, whereas other steps may appear particular to Bayesian methods (e.g. prior predictive checks), but are actually useful for anyone---using Bayesian models or not---to challenge their models of how the world works. Parts of this workflow could be dropped, or expanded as workflows in themselves, given other aims (see Supplement: Which workflow?). For simplicity. many of the smaller but still critical steps are omitted, including visualization, which is required at every step \citep[and for which there are many good resources, e.g.][]{gabryvis}. 

\subsection*{Step 1: Develop your model(s)} 

We start the workflow with what can feel like the biggest step---build a model (or potentially, models) that you want to fit your data to based on your aims. By developing a model designed for your biological question, data and aims, your statistical workflow naturally becomes a scientific workflow. You will more clearly see the assumptions and mechanisms in your model, which is especially valuable given how often our intuition of how models `work' is wrong \citep{kokko2005useful}.  

You likely already have a model, though it may be only verbal or conceptual. For this workflow, however, you’ll need to convert such models into mathematical versions \citep{servedio2014not}. This step is better approached before you collect your data. After data collection, it becomes far more tempting to focus on the particular details in the data and not the latent processes and biological models from which the data were generated. 

Getting to the point where this step is part of your data design and collection, however, requires starting somewhere---with some model (or models) in hand. A suite of resources for `generative' or `narratively generative' modeling can help \citep{statrethink,betangen}, along with two points. First, know that you can and will improve on this skill. Second, as you start, ask lots of questions---and push yourself on your answers---about what you expect and what's reasonable biologically from your model. For example, instead of simply identifying which distribution your observed response variable looks most similar to, ask yourself what ecological processes generate that distribution and how they influence its properties (e.g. its mean, minimum and maximum). Do you expect data below zero? Up to infinity? If not, why not? Effective model building is about efficient brainstorming and this is a critical part of the process. 

As you do this, you'll be generating your model---including its priors, which are important for Bayesian analysis. Assigning priors generally forces you to think about your model with regards to your study system, and interrogate what's probable, possible or actually unreasonable. While many packages (e.g. \textsf{brms, rstanarm}, which fit a suite of pre-defined models) will automatically set default priors, assigning them yourself can quickly disabuse users of their prejudices. For example, you may not think you have a prior on how sunlight affects plant growth, until you realize your `agnostic prior' actually allows plants to grow hundreds of meters per day. 

\subsection*{Step 2: Check your model on simulated data} 

Once you have your model and its priors jotted down, you need to write up your model in a particular modeling language and check it. As with all code: just because it runs, does not mean it does what you think it does. The worst errors often still permit code to run. Whether writing it out in \textsf{Stan}, where you need to be able to write out the full likelihood and set all your own priors, or using a package that writes much of the model for you (e.g. \textsf{rstanarm}), you need a way to verify the code is correct: test data.

\begin{figure}[ht]
\centering
\noindent \includegraphics[width=0.5\textwidth]{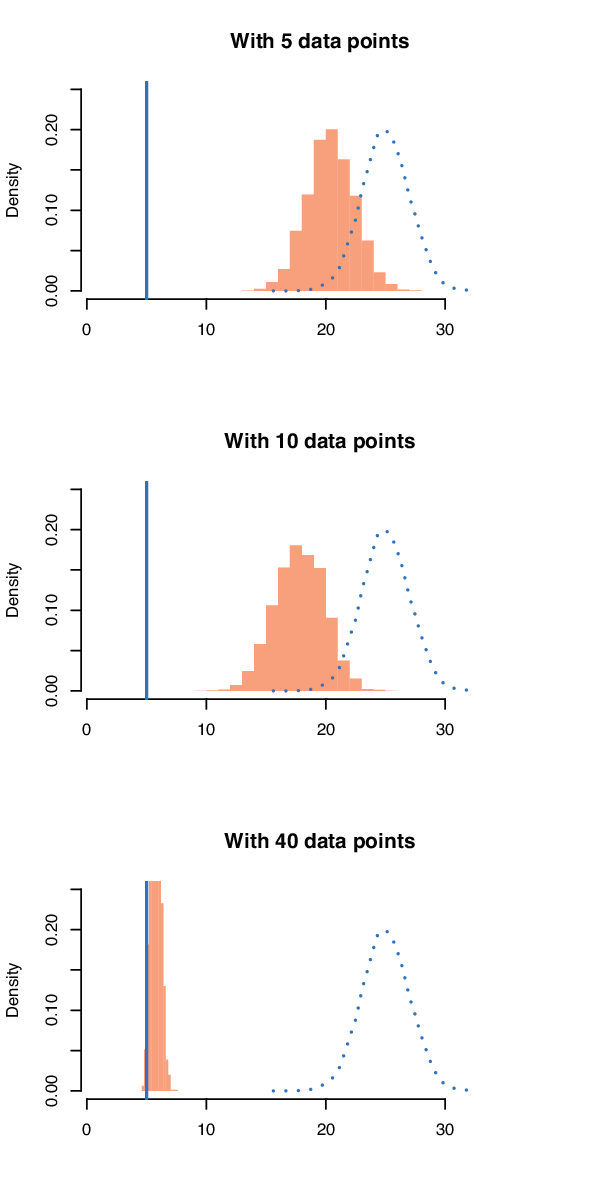}
\caption{A simple example of how to use simulated data to understand calibration issues in a mis-specified model. Here we know the true model underlying the data is $y=\alpha + \text{normal}(0, \sigma)$ where $\alpha$ is 5 (shown as blue vertical line) and $\sigma$ is 2. The model, however, is mis-specified by a prior for $\alpha$ of $\text{normal}(25, 2)$ (dashed blue line), resulting in a posterior (salmon-colored histogram) not centered on the true value. In our experience it is quite rare to have a prior informed by ecological knowledge be so far off, but this is an example. How mis-calibrated the model will be depends on the data: we show examples with a sample size ($N$) of 5, 10 and 40 data points. In practice these studies would allow us to determine how much data we would need to be robust to suspect prior models. }
\label{fig:misspecifyprior}
\end{figure}

Test data (aka `simulated data', or  `fake data,' etc.), and the skills required to build it, are central to this workflow. With `test data' you simulate data from your model in such a way that you can use the resulting data to test if your model code is correct (i.e., you fix values for your model parameters, then test how well your model recovers them, see the Supplement for an example). While there's no guarantee that inferences will always recover the parameter values you set, even when using the correct model, extreme disagreement is often an indicator that something is amiss in the implementation of the model. At the same time these simulation studies can help understand how often a model might lead to the correct inference (see Fig. \ref{fig:misspecifyprior}). As you do this, you will also be calibrating your model---seeing how accurately and precisely it estimates parameters you set and under what conditions. 

This very basic model checking step is uncommon for many ecologists, but critical in our view. If you can simulate data from your model, then you can powerfully---and easily---answer questions related to statistical power, what effect sizes are reasonable, and---most likely---have new insights into how your model suggests the world works, all before looking at any real data. `All models are wrong; some models are useful,' becomes much clearer when you have the power to generate data from your model under any parameter set and sample size you want. Conversely, if you cannot complete this step, you'll struggle to understand if the model fits well, and struggle further to meaningfully interpret the model output, making an apparently simple programmatic task actually encapsulate a far deeper understanding of your model. 

You can learn only so much, however, from data simulated from a particular parameter set. Simulation studies across multiple parameter sets allow you to investigate how robust your inferential performance might be. Prior predictive checks use the Bayesian prior model to set this scope of such simulations. For these, you draw values from your prior distribution (usually randomly in your code) and then explore how your model performs under those draws. Seeing how this influences your resulting output reveals the extent to which your model can capture known variation in your data, and gives insight into whether your model is capable of distinguishing among competing hypotheses. It also serves as a check on the priors you're using (addressing one of the common concerns of those inexperienced with Bayesian models). How exactly to do this depends on your question, model and aims, but many guides can help you think through this \citep{betanprior,wesner2021,winter2023}. \\

 \subsection*{Step 3: Run your model on your empirical data} 
 
The next step is to run the model---you've now validated, test-run and have ready to go---on your exciting new empirical data. Check diagnostics so you know it's running well and adjust until it is \citep[this includes a suite of convergence and efficiency metrics that are well-discussed elsewhere,][]{betanworkflow,gelman2020bayesian,vandeschoot2021,gabryvis}. A model that doesn't converge, or seems to suggest coefficients that are completely at odds with the data, can result from a model that was mis-specified and could never capture real-world variation. For serious problems you may need to reevaluate your model entirely and return to Steps 1 and 2.
 
This is the step many ecologists skip straight to, ourselves included. It's easy to see the appeal: this is the inference step and where you have the opportunity to see new ecological insights. Fitting new data to the model can feel like the moment when you'll learn something new. But, at least in our experience, this is not always the case. When we rush to this step, that first model we fit is often followed by another, and another---perhaps because one does not converge, or the results of another do not make immediate sense. And with the excitement of getting a model to run we can get distracted from what we are actually most interested in---the inference into ecology. 

Following this workflow can make this step much more satisfying. Here the benefits of the workflow may become excitedly apparent: you have estimates in useful units with uncertainty you can understand. You can use this information to draw new conclusions, design new experiments and more---but this is also a point to stop and check your model. 

\subsection*{Step 4: Check your model on data simulated from your empirical model output (also known as posterior retrodictive checks)} 

Once you have your posterior based on your model and new empirical data, it's time to remember that it's wrong, `all models are wrong' after all, and ask how useful it is. You can do some of this through common model-fit diagnostics, such as $R^2$, which compares point predictions to the observed data. With a posterior, however, you can compare an entire distribution of predictions to the observed data. 

\begin{figure}[ht]
\centering
\noindent \includegraphics[width=0.6\textwidth]{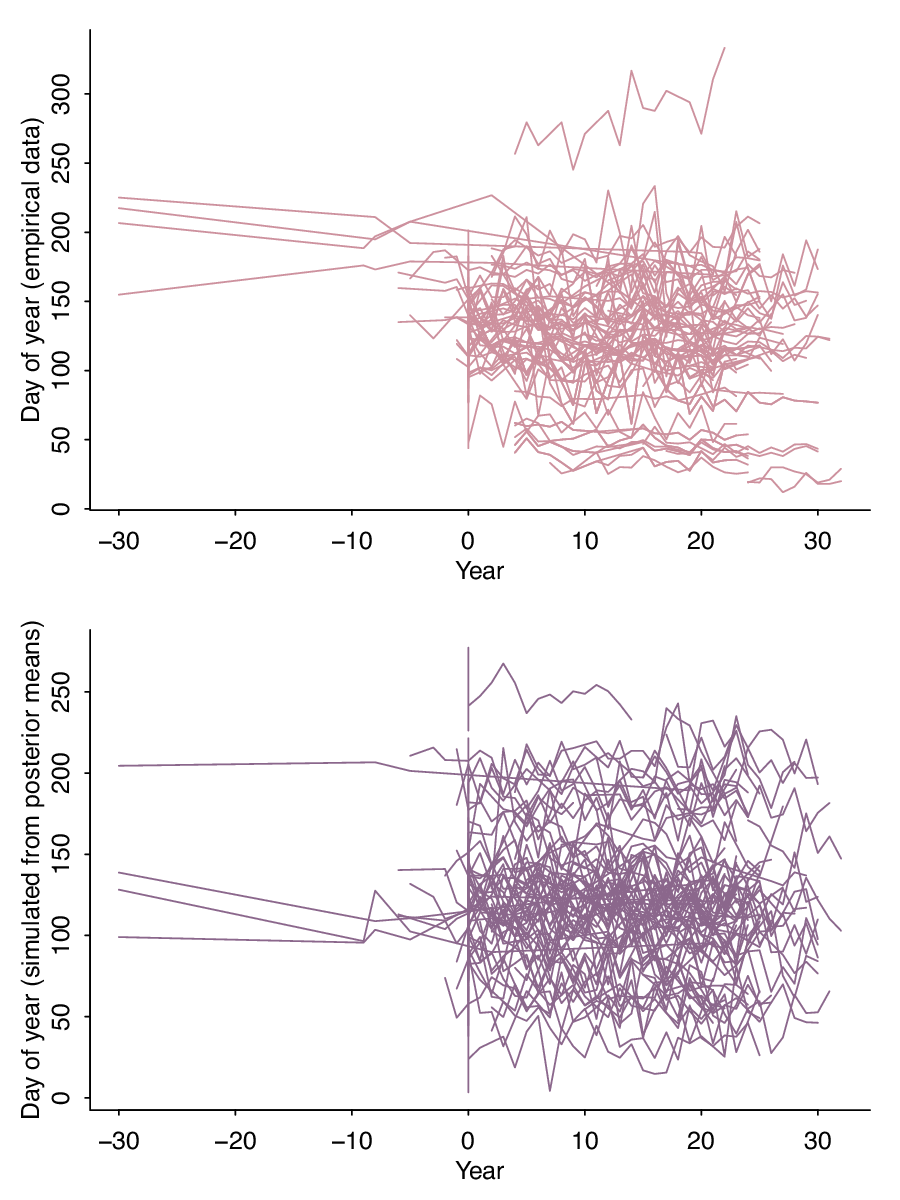}
\caption{Example of a single retrodictive check from time-series data of phenological events over time. The raw data (top, pink) looks similar to one simulated dataset (bottom, purple), based on existing species number, their respective $x$ data, and simulating from the parameters for each species. See `An example workflow' in the Supplement for more details.}
\label{fig:retrodictivecheck}
\end{figure}

This is where simulating from your model can be especially insightful. It will not only indicate that the model isn't adequately fitting the data but also can suggest what the problems might be. Steps 1-2 have set you up well for this, as you have a sense of what different parameter estimates do to the model, and test data provide a sense of how it works on data similar to yours. Now with the parameter estimates from your posterior you can simulate new data from them and see how that new world compares to the observed data---called posterior retrodictive checks (or posterior predictive checks, Fig. \ref{fig:retrodictivecheck}). Exactly how to do these depend---again---on your question, model and aims \citep[see][]{held2010,gelman200ppc,conn2018}. 

Often here you may find big differences from your empirical data, and can start to generate hypotheses for why. For example, you may find patterns that suggest missing grouping factors (e.g. site or biome) through visual posterior retrodictive checks, or you may quickly realize your model predicts impossible numbers for your biological reality. You may begin to see inadequacies in your model, or even potentially your data.  This is one of the main benefits of the workflow: models don't fail silently, they fail with a wealth of context the helps to generate new models and better experiments.

\subsection*{Feedbacks \& workflows}
A key feature of this workflow is that it can be iterated.  If you find that you want to tweak your model then you just return to the beginning, adjust your model, and repeat the rest of the workflow. In this way, fitting multiple models is encouraged, but this is distinct from the quest for a minimum adequate model or one `best' fit. Feedbacks in this workflow are focused far more on what is biologically reasonable, and understanding the utility---and limits---of inference from your data for your model.  And there are big benefits to it. 

This process more fully integrates mathematical modeling into statistical modeling. To complete Steps 1-2, you have to understand the underlying math of your model enough to simulate data from it. This can be challenging at first (e.g. recalling how to simulate $y$ data for a simple linear regression is not straightforward when you rarely do it), but is immensely beneficial to forcing you to understand your model and its consequences. Indeed, we have found the greatest insights come not from the step we all know best---fitting the model with empirical data (Step 3)---but from every other step in this workflow. 

\section*{How this workflow changed our science} 

As we have used this workflow, how we approach our statistical models has changed. These changes have generally been similar for each of us. We suspect they are not unique to us, our study systems, or our questions. Instead, we think they represent common approaches to statistical modeling that could help ecology advance, much as we believe they have helped our science advance. 

 \subsection*{Looking at parameters, not $p$-values} 
Before this workflow, not all of us commonly discussed the values that parameters in our model took---things like the slope and intercept (two common model parameters) were sometimes reported, but we did not know them as well as we knew whether the $p$-value for the slope was $<0.05$. This changes quickly when you need to build simulated data (Step 2); for example, when modeling phenological events (observations of biological events on numbered days within the calendar year: 1-365 most years) it is not uncommon to find seemingly-reasonable models generating predictions of events on the non-existent calendar day of 1000. This focus on the value of parameters scales up through this and other modeling workflows. Having a better sense of parameter values across different biological contexts, model parameterizations, and time periods gives a better sense of how the biological world works, including what's reasonable, possible or wildly unrealistic. 

 \subsection*{Thinking about priors} 
This workflow also can quickly ease the common concern of those unfamiliar with Bayesian approaches: priors. Often treated as the big bad wolf of Bayesian, or the unseen hand producing biased model fits, according to some. In reality, how much priors influence your model fit is up to your model and your data. Depending on those two parts, the likelihood (influenced by your data) can easily overwhelm your priors (Fig. \ref{fig:misspecifyprior}). Indeed, most work on the dangers of priors and `prior misspecification'  focuses on cases where you have very little data for the model you're trying to fit. Priors, however, can only matter more than you know when you fail to think through and check them---that is, you skip Steps 1-2. 

\subsection*{Understanding nonidentifiability}
This workflow also taught us a lot about identifiability, which refers to when all parameters in a model can be uniquely identified with infinite data. More common in our experience is nonidentifiability. Models can be nonidentifiable in several ways, including when mathematically some parameters cannot be distinguished from each other no matter how much data we have. 
A statistical kin of non-identifiability is degeneracy.  Degeneracy concerns the kinds of complex uncertainties that can arise from finite data sets \citep{gelmanhill}, and something we have often found in Steps 2-3 of our workflow.

Nonidentifiability and degeneracy can come up in many ways in ecology, and be hard to see if you rush through model fitting. But if you have to write out your model and simulate data (Steps 1-2), you may suddenly realize lots of places for nonidentifiability and degenercies to live. For example, when species do not occur across most sites, a model including separate parameters for site and species is often degenerate, but there's often no warning in packages to tell you this. We have become far better at noticing nonidentifiability and degenercies based on this workflow---and we have adjusted how we collect data and interpret results because of it. 

 \subsection*{Knowing our limits}  
Once we noticed how pervasive non-identifiability and degeneracies are how our approach to building complex models also changed.
Before using this workflow, we might start with more complex models, then simplify them until their fits converged in our given software package. Often these were hierarchical models with many levels---for example, including every column of site, plot, transect and quadrat in our dataframe, without stopping to check how well sampled they were, or what degeneracies they might introduce. 

We also often fit a suite of interactions: multiple two-way interactions and the occasional three-way interaction were common fare. But in simulating data, and fitting models to real, messy, imbalanced data using the workflow we came to see how much we were asking of our data and models together. Fitting a two-way interaction with half the effect size of a main effect takes a 16X sample size, compared to fitting the main effects alone \citep[the main effects then average over the interactions, see][for more details]{regotherstories}. This is sobering. It's more sobering when you see it played out again and again through this workflow. 

We now both add complexity and simplify based on a more careful reckoning. Often our starting model is not simple.  It often includes grouping factors that may be difficult to fit, but only those grouping factors that we see as absolutely critical to the question, model and data at hand. We still add and consider additional grouping factors and interactions, but we do so with a careful idea of how stable the model given the data likely is with them, and we rarely fit complex three-way interactions or similar---unless we have carefully designed the model and data collection for that aim. 

 \subsection*{Fitting bespoke models} 
Once we have identified the limits to our data, we can fit a bespoke model to estimate the parameters we are actually interested in rather than the numbers that are convenient to estimate. Models can be specifically designed to estimate and report effects in relevant units (e.g. per degree C of warming)---always with estimated uncertainty. Such flexibility is incredibly powerful in ecology where data are often influenced by complex spatial or temporal patterns, non-linear processes are widespread, and common data types are non-Gaussian (e.g. counts, percent cover, etc.). 

Bayesian models have many benefits, but an often-mentioned one is that `you can fit any model you want.' While this is not entirely true \citep{BDA,reid2019}, Bayesian modeling options can feel limitless when compared to the models ecologists can fit in popular modeling packages (e.g. \textsf{lme4} package in \textsf{R}). As long as you can write out the likelihood of your desired model \citep[and sometimes even if you can't,][]{Sunnaaker2013} and assign priors to all parameters, you can generally `fit' the model. However, the flexibility of Bayesian models is sometimes seen as a weakness: you can fit almost whatever you want, but critical parts of your model might be almost entirely unimpacted by your data. In ecological model fitting, we're currently most often interested in parameter estimates strongly informed by our data, making this problem sound especially dangerous. In reality, however, this problem is not related to modeling, but to experimental design---and a flawed experimental design leading to low power for your model is much easier to see through using our workflow compared to using traditional null hypothesis testing methods. Low power usually becomes very obvious in Step 2 when you need to recover your parameter estimates and find you need simulated data very different (e.g. higher sample size, lower error) than your empirical data to achieve this.

\section*{How this workflow intersects with ecological training} 

This four-step workflow is a simplified version of the current best practices for Bayesian model fitting  \citep{betanworkflow,vandeschoot2021}, but many of the skills required are not part of traditional ecological training. Writing out the math behind most statistical models enough to complete Steps 1-2 bleeds into the skillset usually reserved for those working on theory, where coding and simulating from a model are common tasks. In contrast field, lab and otherwise empirical-data based ecologists often fit models they could not simulate data from. This dichotomy seems short-sighted in our current era of bigger, messier data and greater computational methods poised to handle such messy data. Further, the increasingly computational toolkit of the modern ecologist makes it easier to bridge the gap between ecological models and their underlying math. 

A reasonably competent coder could easily simulate data under a complex model that they might not have the mathematical expertise to solve analytically---if doing so was part of their training and the workflows they regularly use. 
Simulating data rapidly clarifies underlying assumptions. While training in frequentist methods often includes memorizing assumptions for a particular test, or steps specifically designed to test particular assumptions (e.g. normal quantile plots), this workflow requires no such training. Instead it requires only the skills to identify whatever the assumptions have been encoded in your models. 

Uptake of new Bayesian \textsf{R} packages highlight that Bayesian methods are no longer the purview of only a few, and these changes come alongside advances in Bayesian workflows, algorithms, and visualizations \citep[e.g.][]{betanworkflow,vandeschoot2021,gabryvis}. Ecology must adapt its training to embrace these shifts. While this is an active area, we highlight three major changes.

(1) Prior `beliefs' are changing. Best practices for determining priors is an active area of statistical research \citep{BDA,regotherstories,betanprior}, and training should reflect current best practices. These include that `non-informative priors' are a misnomer, as they are often informative  \citep{lemoine2019}, and priors can easily be `weakly informative.' Thus a strong focus on the dangers of priors in training can be overkill, and verges on scaremongering. 

Current training often includes a very strong focus on mathematically-convenient priors, because of the importance of conjugate priors in closed form solutions for particular posteriors. Modern algorithms, such as HMC, do not require conjugate priors. Prior predictive checks provide a far more powerful way to understand how priors work within a particular model, and are more useful than rules about which priors should be fit in certain cases or memorizing which priors are conjugate \citep{betanprior}. 

(2) `Random effects' are not just random. Hierarchical models contain grouping factors, sometimes referred to as random effects, such as species or individual. The term `random effects', however, is misleading, imprecise and thus no longer recommended \citep{gelmanhill}. In ecology, it also carries with it many older `rules' of what is `random' versus `fixed,' including that `random effects are things you don't care about' (for example the `block' effect from a randomized block design). Training in posterior retrodictive checks (Step 4) may reshape these views, as hierarchical effects are (by definition) drawn from an underlying distribution---meaning they can predict outside of the specific set  sampled (for example, to predict for a new species or individual), whereas the same is not true for most categorical `fixed' effects.

(3) P-values, and null hypothesis testing in general, are easily misleading, and there’s no easy fix for that. The replication crisis, rampant in other fields, is based in part on an overly hopeful belief that $p$-values will separate the signal from the noise, with one easy number. In reality, small sample sizes, lack of routine reporting of interpretable effect sizes, fitting of many models without adequate explanation, poor data and code reporting habits all increase the chance of finding `significance' at a level of $\le0.05$ \citep{halsey2015,loken2017}. This reality means a similar crisis is likely lurking in ecology, especially given small sample sizes alongside a tendency to fit complicated models with multiple interactions. The answer to this, however, is not to make $p$-values smaller \citep{halsey2015,colquhoun2017}, nor is it Bayesian approaches. 
The answer is training in workflows designed for careful model building, model fitting and model interrogation informed by ecological theory and understanding---including the one we outline here. 

\emph{Acknowledgements:} Comments from F. Baumgarten, D. Loughnan and N. Pates improved this manuscript and J. Ngo improved Figure 1. 

\iffalse
\subsection*{Conclusions}
\begin{enumerate}
\item Ecologists cannot simulate their stats (or simple systems for that matter). Evolutionary biologists can. (And the field is better for it.)
\item Maybe hint at that you need these skills (and unit testing) given rise of AI?
\end{enumerate}

{\bf Take home messages (maybe)}
\begin{enumerate}
\item You should not fit a model you cannot simulate
\item Fit simpler models
\item Know your nonidentifiability
\end{enumerate}
\fi

\clearpage

\section*{Tables \& Figures}

\begin{table}
\caption{A set of major terms used with their simplified definitions.}
\begin{tabular}{ p{3 cm}  p{13 cm} }  \hline \hline
 \emph{Term}   & \emph{Definition}\\ 
\hline \hline
calibration & analyzing how often an estimate is close to the true value over an ensemble of hypothetical observations.  An exact calibration would requires simulating from the true data generating process which is impossible in practice.  We can, however, calibrate to data simulated from the configurations of models we plan use to fit to our data (\emph{Steps 1-2}) so we understand the models better, including their limits given data similar to ours. We emphasize simulations to calibrate model behaviors consistent with our ecological systems and understanding (e.g. working within a limited set of parameter ranges through prior predictive checks). In contrast to this approach, frequentist method are calibrated against all possible behaviors, which is not only impractical for complicated models it’s also irrelevant given that the most extreme behaviors are unlikely to manifest in reality. \\\hline
degeneracy & complex uncertainties that come from a mix of sources, including, non-identified models and cases where the data cannot well inform model parameters. When the data are not informing the parameters that we care about, this highlights a measurement issue. Identifying these problems in simulation studies can highlight when we need a better experimental design (e.g. sampling for more overlapping species across sites, or changing what we measure, etc.).  \\\hline
non-identifiability & when all parameters in a model cannot be uniquely identified with infinite data \\\hline
prior & an distribution of reasonable values for a parameter based on fundamental biological and ecological understanding, previous research, or other sources \\\hline
statistical model & Mathematical approximations of the true data generating process labeled with numerical parameters.  Evaluating a statistical model on observed data gives a likelihood function that quantifies how compatible different parameters are with the observed data, and hence can be used to `fit' the best parameters. In this article, we often simplify to `model.' See also the Supplement: What's a model? \\\hline
posterior & product of the likelihood and prior; that is, a probability distribution that quantifies how compatible different model parameters are with both the observed data and the domain expertise encoded in the prior model. \\\hline
workflow & a set of steps to achieve a goal, with those steps designed to help organize the process, and ideally make it more systematic  \\\hline
\hline
\end{tabular}
\label{tab:glossary}
\end{table}

\newpage

\end{document}